\documentclass[12pt,preprint]{aastex}
\usepackage{amsmath}
\usepackage{natbib}

\begin{document}

\bibliographystyle{apj}

\title{The Atmospheric Chemistry of GJ 1214b: Photochemistry and Clouds}

\author{Eliza Miller-Ricci Kempton}

\affil{Department of Astronomy and Astrophysics, University of California, 
      Santa Cruz, CA 95064}

\email{ekempton@ucolick.org}

\author{Kevin Zahnle}

\affil{NASA Ames Research Center, Moffett Field, CA 94035}

\author{Jonathan J. Fortney}

\affil{Department of Astronomy and Astrophysics, University of California, 
      Santa Cruz, CA 95064}

\begin{abstract}

Recent observations of the transiting super-Earth GJ 1214b reveal that its 
atmosphere may be hydrogen-rich or water-rich in nature, with clouds or hazes
potentially affecting its transmission spectrum in the optical and 
very-near-IR.  Here we further examine the possibility that GJ 1214b does 
indeed possess a hydrogen-dominated atmosphere, which is the hypothesis that is
favored by models of the bulk composition of the planet.  We study the effects 
of non-equilibrium chemistry (photochemistry, thermal chemistry, and mixing) on
the planet's transmission spectrum.  We furthermore examine the possibility 
that clouds could play a significant role in attenuating GJ 1214b's 
transmission spectrum at short wavelengths.  We find that non-equilibrium 
chemistry can have a large effect on the overall chemical composition of GJ 
1214b's atmosphere, however these changes mostly take place above the height
in the atmosphere that is probed by transmission spectroscopy.  The effects
of non-equilibrium chemistry on GJ 1214b's transmission spectrum
are therefore minimal, with the largest effects taking place if the planet's
atmosphere has super-solar metallicity and a low rate of vertical mixing.  
Interestingly, we find that the best fit to the observations of GJ 1214b's 
atmosphere in transmission occur if the planet's atmosphere is deficient in
CH$_4$, and possesses a cloud layer at a pressure of $\sim$200 mbar.  This is 
consistent with a picture of efficient methane photolysis, accompanied by 
formation of organic haze that obscures the lower atmosphere of GJ 1214b at 
optical wavelengths.  However, for methane to be absent from GJ 1214b's 
transmission spectrum, UV photolysis of this molecule must be efficient at 
pressures of greater than $\sim 1$ mbar, whereas we find that methane only 
photolyzes to pressures less than 0.1 mbar, even under the most optimistic 
assumptions.  An alternative explanation of the observations of GJ 1214b is 
that the atmosphere is water-rich, although this interpretation conflicts with 
the findings of Croll et al.~(2011), who measure a low mean molecular weight 
for the planet's atmosphere.  Additional observations at wavelengths 
corresponding to mid-IR water and methane features in GJ 1214b's transmission 
spectrum should break the degeneracy between the two possible cases.

\end{abstract}


\section{Introduction \label{intro}}

The transiting super-Earth GJ 1214b is ideally suited to follow-up observations
aimed at characterizing the planet's atmosphere due to the fact that it orbits 
a relatively bright M-star ($I = 11.52$), producing a deep transit with an 
observed depth of 1.37\% \citep{ber11, cha09}.  The planet additionally 
transits quite frequently with its 1.58 day orbital period (a = 0.0143 AU), 
which further facilitates observational follow-up.  Recently, the first 
observations of GJ 1214b's atmosphere were reported by \citet{bea10}.  These
observations, taken with the FORS2 instrument on the VLT reveal a flat 
transmission spectrum in the near-IR from 780 to 1000 nm. 

GJ 1214b has an observed mass and radius of 6.55$\pm$0.98 $M_{\earth}$ and 
2.68$\pm 0.13$ $R_{\earth}$ \citep{cha09} respectively, implying a low density 
(1.88 g/cm$^3$) that requires the planet to possess a significant atmosphere 
\citep{net10, rog10}.  Models of the interior of GJ 1214b suggest that the 
planet either possesses a thick hydrogen atmosphere comprising anywhere between
0.05 and 2.5\% of the total mass of the planet, or alternatively that GJ 1214b 
is a water-rich planet with a thick steam atmosphere.  Under the water-rich 
scenario \citet{rog10} find that the planet must be at least 60\% water by mass
to reproduce the planet's observed radius to within its 2-$\sigma$ error bar.
\citet{net10} find that a similarly large water-to-rock ratio of greater than 
6:1 is needed to reproduce the planet's radius  at its best-fit value.  
However, a bulk planetary composition of such high abundances of water is 
thought to be improbable from a formation standpoint.  Intermediate atmospheres
between these two end member cases composed of a combination of water vapor and
hydrogen gas are also consistent with the planet's observed mass and radius, 
and these would require a smaller fraction of water to reproduce the planet's 
bulk density.  Differentiating between the two proposed classes of atmospheres 
(H$_2$-rich and H$_2$O-rich) has important implications for understanding the 
formation history of GJ 1214b and other similar super-Earths.  A hydrogen-rich 
atmosphere for this planet implies that GJ 1214b acquired its atmosphere either
by accretion of nebular gasses or from outgassing of significant amounts of 
hydrogen as the planet cooled and solidified from a magma ocean state, whereas 
a water-rich planetary (and atmospheric) composition implies that the planet 
formed from ice-rich material beyond the snow line before migrating in to its 
current location. 

Breaking the degeneracy between the two classes of proposed atmospheres is not
possible through refined measurements of GJ 1214b's mass and radius, due to
significant degeneracies in the mass-radius relationship for solid exoplanets
\citep{val06, for07, sea07, sot07}.  Instead, observations of the planet's 
atmosphere that constrain its composition are necessary to break the degeneracy
between an atmosphere composed of either predominantly hydrogen or water vapor.
The most straightforward way to ascertain the composition of GJ 1214b's 
atmosphere is to observe the planet's atmosphere in transmission, owing to the 
relationship between the amplitude of spectral features in transmission and 
the mean molecular weight of the planet's atmosphere \citep{mil09}.  This
relationship results from the fact that the depth of transmission spectral
features is directly proportional to the atmospheric scale height, which
in turn varies inversely with the mean molecular weight.  For GJ 1214b, a 
hydrogen-rich atmosphere should produce spectral features in transmission with 
amplitudes of 0.1-0.3\% relative to light from the host star, whereas the 
signature of a water vapor atmosphere would only be present at the 0.01\% 
level and would be undetectable with most current instrumentation used for 
exoplanet characterization \citep{mil10}.

\citet{bea10} observe a flat transmission spectrum for GJ 1214b between 780 and
1000 nm, which is consistent with a small scale height, high mean molecular 
weight atmosphere composed of at least 20\% H$_2$O by volume.  An alternative 
interpretation of the data is that GJ 1214b's atmosphere is hydrogen-dominated,
but high-altitude clouds or hazes obscure the deeper regions of the planet's 
atmosphere where molecular absorption features are predicted to originate in 
transmission over the wavelength range of the \citet{bea10} observations.  
(Here we define clouds as being formed by equilibrium condensation processes,
whereas a haze results from the action of photochemistry.) In the case of the
\citet{bea10} transmission observations, the cloud layer is required to become 
optically thick at heights corresponding to pressures of less than 200 mbar, in
order to be consistent with the data at the 1-$\sigma$ level.  Without a cloud
layer, transmission observations would probe even higher pressures at the 
wavelengths of the \citet{bea10} observations, due to the low opacities 
predicted for this spectral range.  Because scattering from cloud or haze 
particles should be more efficient at short wavelengths, clouds will have a 
smaller effect on the transmission spectrum at longer wavelengths in the IR 
where deep molecular absorption bands should still be present -- whereas a 
water-rich atmosphere would produce a flat spectrum at the $\sim 0.1$\% level 
across all wavelengths.  

For this reason, additional observations taken at infrared wavelengths could 
serve to confirm or rule out the hydrogen-rich nature of GJ 1214b's atmosphere.
Indeed, \citet{cro11} find that GJ 1214b produces a deeper transit in Ks-band 
($\sim$2.15 $\mu$m) than in J-band ($\sim$1.25 $\mu$m) at the 3-$\sigma$ level,
which is consistent with the predicted signature of a hydrogen-rich atmosphere.
However, these results have been difficult to rectify with longer wavelength 
data at 3.6 and 4.5 $\mu$m from \citet{des11}, who also find a flat 
transmission spectrum that is consistent with the \citet{bea10} transit depth. 
This is an issue that we attempt to address in this paper.

Another possibility that has yet to be explored is that disequilibrium 
chemistry may play a significant role in determining GJ 1214b's transmission 
spectrum.  Previous modeling efforts have only considered hydrogen-rich 
compositions of GJ 1214b's atmosphere in thermochemical equilibrium with 
metallicities between 1 and 50 times solar \citep{mil10} as a starting point, 
but this may not be a valid assumption for the actual state of the planet's 
atmosphere.  Non-equilibrium chemistry is known to play a strong role in 
determining the atmospheric composition for many of the planets and 
(planet-sized moons) in our solar system and is also responsible for the 
formation of clouds or hazes in the atmospheres of certain solar system bodies 
(e.g.~H$_2$SO$_4$ clouds on Venus and hydrocarbon hazes on Titan).  Processes 
that can perturb a planetary atmosphere away from a state of thermochemical 
equilibrium include photochemistry, dynamical mixing and winds, and are 
ultimately limited by the finite amount of time required for chemical reactions
to proceed.  Overall, if the timescale for mixing or photochemical destruction 
of a species is shorter than the timescale for chemical reactions to return the
gas ensemble to its equilibrium state, then non-equilibrium chemistry cannot be
ignored.  

In this paper we study the effects that non-equilibrium chemistry can have on
the overall composition of GJ 1214b's atmosphere, assuming that it is composed
of hydrogen-rich gas.  We apply a photochemical model, which accounts for
UV photolysis of molecules, vertical mixing in the atmosphere, and the finite
timescales for chemical reactions to proceed.  The result is a suite of 
non-equilibrium chemical compositions for GJ 1214b's atmosphere, based on a 
grid of assumptions for the planet's metallicity and eddy mixing rate, and UV 
irradiation by the host star.  For the non-equilibrium chemical compositions 
that we compute, we determine the transmission spectra that would result. 
We then compare the resulting non-equilibrium transmission spectra against the 
currently available data for GJ 1214b.  We also delve further into the question
of clouds or hazes in GJ 1214b's atmosphere as a possible explanation for the 
planet's flat transmission spectrum at near-IR wavelengths.  The paper is laid 
out as follows.  In Section~\ref{methods} we describe our methodology and 
modeling efforts.  In Section~\ref{results} we present our modeling results for
non-equilibrium chemical abundances and the effects on the transmission 
spectrum of GJ 1214b, along with constraints on clouds in GJ 1214b's 
atmosphere.  In Section~\ref{obs} we explore the extent to which our 
non-equilibrium transmission spectra can reproduce the observations of GJ 1214b
in transmission.  We conclude with some closing remarks and discussion in
Section~\ref{concl}.  

\section{Methodology \label{methods}}

Most previous efforts to model the atmospheric structure and spectra of 
transiting exoplanets have relied on the assumption that the planetary
atmosphere resides in a state of thermochemical equilibrium 
\citep[e.g.][]{sea00, for05, mil09}.  This assumption holds if the atmospheric 
temperature is sufficiently high, the planet exists in an isolated environment,
and vigorous atmospheric mixing does not occur.  However, most transiting 
extrasolar planets orbit in exceptionally close proximity to their host stars, 
and they therefore sit in environments that receive high levels of UV 
irradiation.  To first order, UV photolysis is expected to affect the 
chemistry of the upper atmosphere of these planets, forcing the atmospheric 
chemistry away from a state of equilibrium.  Vertical mixing and winds can 
additionally mitigate or exacerbate these effects.  Already, observational 
signatures of non-equilibrium chemistry have been observed for several 
transiting hot Jupiters \citep[e.g.~][]{swa08, swa09b, swa09, mad11, mad11b}, 
and it is becoming increasingly clear that the effects of photochemistry on 
exoplanet atmospheres cannot be entirely ignored.  
 
A number of authors have recently begun researching the effects of 
non-equilibrium chemistry on the atmospheres of highly irradiated extrasolar
planets.  A brief description of this work is as follows.  Early work was 
performed by \citet{lia04}, who looked at the effects of hydrocarbon chemistry 
on the atmospheres of close-in giant planets. More recently, \citet{zah09} and 
\citet{zah11} applied a photochemical model to look at the effects of sulfur 
and hydrocarbon chemistry, respectively, on hot Jupiter atmospheres using the 
same model that we have adopted for this paper.  Work by \citet{lin10} 
additionally studied the effects of photochemistry specifically on the 
atmosphere of the close-in extrasolar giant planet HD 189733b.  \citet{mos11} 
performed the first study to not only look at atmospheric chemistry but also to
predict the observational signatures of disequilibrium processes in the 
atmospheres of transiting exoplanets by modeling transmission and emission 
spectra for HD 209458b and HD 189733b with the abundance profiles from their 
photochemical modeling.  \citet{sha11} additionally looked at the effects of
non-equilibrium chemistry on the transmission spectrum of the hot Neptune GJ 
436b, using the results from the \citet{zah11} photochemical models.  In a 
reverse approach \citet{mad09}, \citet{mad11}, and \citet{mad11b} have 
attempted to back out the chemical abundances of certain key species in hot 
Jupiter atmospheres, based on the appearance of the planets' transmission and 
emission spectra without using any assumptions of chemical equilibrium.  This 
backward modeling approach will ultimately allow for detailed fitting of data 
to simultaneously determine atmospheric structure and abundance profiles for a 
given exoplanet atmosphere.  However, the fitting procedure is currently 
challenged by the fact that the models are under-constrained by the available 
data.  Backward modeling will become particularly useful in the future once 
full spectra of exoplanet atmospheres become available.  In the meantime, given
the minimal quantity of available data for GJ 1214b, we use a forward modeling 
approach in this paper, similar to the work of \citet{mos11}, to predict the 
abundance profiles and spectral signature for this transiting super-Earth. 

\subsection{Photochemical Model}

We calculate atmospheric abundances for the transiting super-Earth GJ 1214b 
that result from non-equilibrium chemical processes (photolysis and vertical 
mixing) using the photochemistry model developed in \citet{zah09} and 
\citet{zah11} for hot Jupiter exoplanets.  This model is based on the 1-D 
photochemical kinetics code initially described in \citet{kas89} and 
\citet{kas90}.  While GJ 1214b is not a hot Jupiter, this code is generally 
applicable to exoplanets with hydrogen-rich atmospheres.  Since \citet{zah11} 
the code has been completely re-written in the C programming language.  Updates
have been made to further generalize the code for calculating the composition 
of any given exoplanet atmosphere.  Specific improvements include functionality
for using an arbitrary temperature-pressure profile and increased flexibility 
in the altitude step size, along with a generalized treatment of UV photolysis.
We use the photochemical kinetics code to calculate chemical abundances for 61 
atomic and molecular species in GJ 1214b's atmosphere as a function of height 
(parametrized within the code in terms of atmospheric pressure) by 
simultaneously solving the equations of continuity and flux.  In this study, 
we solve for 698 chemical reactions including 33 photolysis reactions.  The 
reaction rates and photolysis cross sections employed in this work are listed 
in \citet[][their Tables 2 and 3]{zah11}.  Reverse reaction rates are 
calculated using the method outlined in \citet{vis11} as outlined below.

For reactions of the form ${\rm A} + {\rm B} \rightarrow {\rm C} + {\rm D}$ 
with forward reaction rate $k_f$, the reverse rate (i.e., the rate for 
$ {\rm C} + {\rm D} \rightarrow {\rm A} + {\rm B}$) is 
$k_r=k_f\exp{\left(-\Delta G/RT\right)}$, where $\Delta G$, the Gibbs free 
energy, is obtained from enthalpies and entropies of the of the reactants and 
products, $\Delta G=H_A+H_B-H_C-H_D - T\left(S_A+S_B-S_C-S_D\right)$.
Associative reactions of the form ${\rm A} + {\rm B} \rightarrow {\rm AB}$, the
rate for the reverse reaction (dissociation of AB) is 
$k_r=k_f\left(kT/P_{\circ}\right) \exp{\left(-\Delta G/RT\right)}$, where 
$P_{\circ}=10^6$ dynes/cm$^2$ is one atmosphere.  This must be done in both the
low pressure and high pressure limits.  Similarly, the associative reverse of a
dissociative reaction is given by 
$k_r=k_f\left(P_{\circ}/kT\right) \exp{\left(-\Delta G/RT\right)}$.
For this code we have fit the reverse reaction rates to the standard Arrhenius 
form $k=AT^b\exp{\left(-T/C\right)}$.  This maximizes code flexibility and 
exploits the thermodynamic limit as an error check.  The fits can be quite 
good, although the asymptotic match to the thermodynamic equilibrium limit is 
imperfect. 

For all of our calculations we employ an underlying temperature-pressure (T-P)
profile for GJ 1214b's atmosphere that was presented in \citet{mil10} for 
solar composition gas in chemical equilibrium and assuming planet-wide 
redistribution of heat (shown also in Figure~\ref{f7}).  The effective 
temperature and surface gravity for this model are $T_{eff} = 555$ K and 
$g = 8.95$ m/s respectively.  We do not include feedback on the T-P profile 
based on the changes in atmospheric composition that result from the 
photochemical modeling.

Due to the fact that the vertical eddy mixing rate and the atmospheric 
metallicity are unknown for GJ 1214b, we generate a grid of models in both of 
these parameters.  For the eddy diffusion coefficient we calculate models with 
constant mixing rates of $K_{zz} = 10^6$, $10^7$, and $10^9$ cm$^2$/s.  For 
metallicity, we choose values ranging from solar composition to metal enhanced
at 1, 5, and 30 times solar.  For all of our calculations we use a stellar 
zenith angle of $\theta = 30 ^{\circ}$.  We have done some additional tests
with higher values of $\theta$ corresponding to longitudes closer to the
terminator, and we have found the effects to be minimal, although a slightly
lower level of photodissociation is found.

As a lower boundary condition, we assume that GJ 1214b resides in 
a state of chemical equilibrium at the base of the atmosphere.  Generally
speaking, deep in the atmosphere of a planet at sufficiently high temperatures 
and pressures the chemistry should approach thermal equilibrium.  However, this
may not be valid if the atmosphere is thin, in which case interactions between 
the planet's surface and its atmosphere should be taken into account.  For GJ 
1214b, the minimum mass of its hydrogen envelope corresponds to an atmosphere 
of at least 450 bars, and the temperature at the base the atmosphere should 
exceed 1175 K based on calculations of the T-P profile.  We calculate 
equilibrium abundances at the base of the atmosphere using
the Gibbs free energy minimization code described in \citet{mil09}, which is in
turn based on the method outlined in \citet{whi58}.  Some of the 61 species 
considered in the photochemistry code are not included in the equilibrium 
chemistry code because they are not predicted to be present in any significant 
quantities in equilibrium.  For these molecules, we set their initial 
abundances at the bottom of the atmosphere to an arbitrary low value of 
10$^{-40}$.  We place our lower boundary for all of our models at a pressure of
1000 bars, where the 
atmospheric temperature is predicted to be $T \approx 1400$ K from the T-P 
profile that we employ in this work.  We have performed limited tests to 
determine the sensitivity of our results to the pressure that we choose as the 
bottom boundary of the atmosphere, and we do not find any strong dependencies 
for pressures greater than 100 bar.  We have additionally assured that our 
lower boundary lies below the height where CO-CH$_4$ quenching takes place 
(around 100 bar).  Below this height we find that the abundance profiles for 
the carbon-bearing species follow their equilibrium values.  We do not 
attain quenched abundances for N$_2$ and NH$_3$ even at the 1000 bar lower 
boundary of our atmosphere, but we show in subsequent sections that this does 
not have a strong effect on our results.   

As an upper boundary condition we set a 
zero flux lid at 1 $\mu$bar with no flow allowed into or out of the top of the 
atmosphere.  From theoretical calculations, GJ 1214b is thought to be 
potentially undergoing atmospheric mass loss at a rate of $\sim 9 \times 10^8$ 
g/s \citep{cha09}, however this rate is unconstrained by observations and we 
therefore choose to ignore the effects of mass loss in our photochemical 
calculations.  

\begin{figure}
%\begin{center}
%\includegraphics[scale=0.81]{f1.eps}
%\end{center}
\plotone{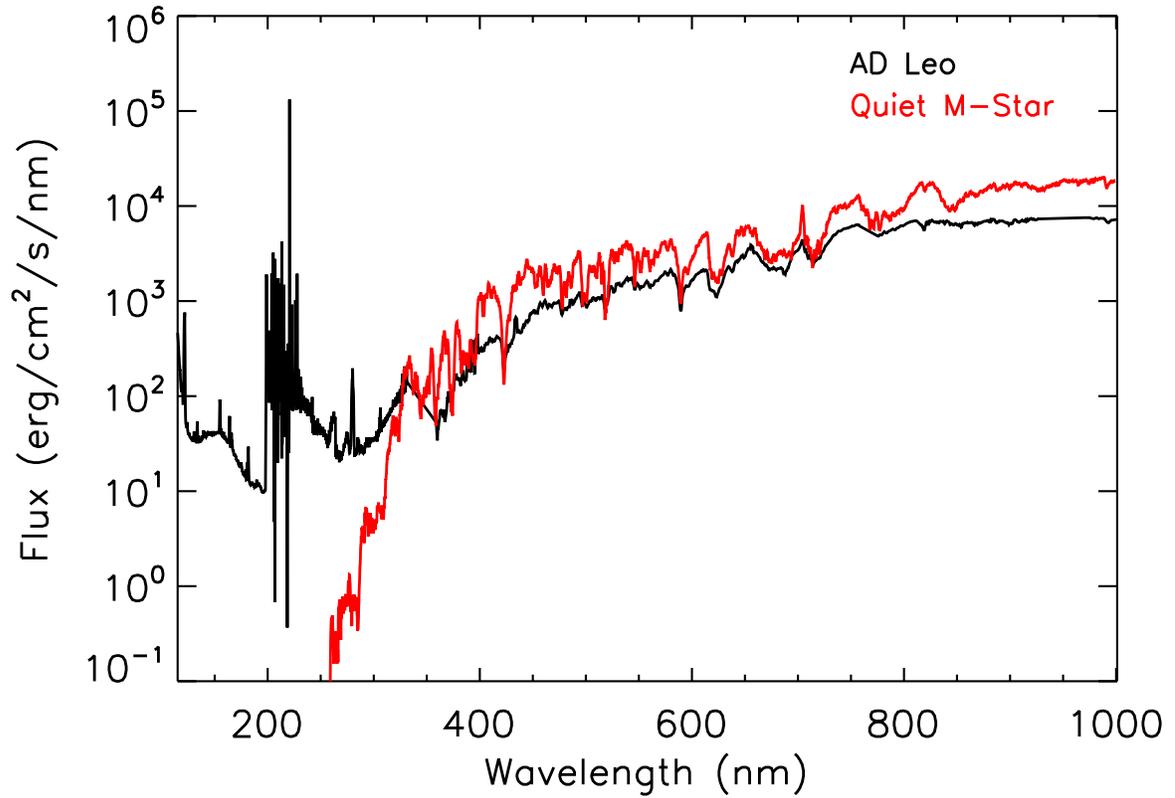}
\caption{Spectra for the two stars that we use in our photochemical modeling.
        The red line is a modeled spectrum for a quiet M-star from 
	\citet{hau99}.  The black line is the time-averaged spectrum of the 
	M4.5V star AD Leo from \citet{seg05}.  The fluxes listed on the y-axis
	are the incident fluxes at the orbital distance of GJ 1214b (a = 0.0143
	AU).
        \label{f1}}
\end{figure}

To calculate photolysis rates it is necessary to know the amount of stellar UV 
flux that is absorbed by the planet's atmosphere.  However, the UV spectrum
of GJ 1214 has not been measured.  Generally, M-stars such as GJ 1214 tend to 
be relatively more active than early-type stars, with stellar activity 
decreasing as a function of stellar age.  In theory, we can place GJ 
1214 on a stellar type vs.~age diagram \citep[such as from][]{sel07} to 
determine its expected level of UV flux.  However, given the large uncertainty 
in the star's age \citep[6 $\pm ^4 _3$ Gyr;][]{cha09}, the constraints that we 
can place on its UV flux are essentially meaningless.  For this reason, we 
choose to include UV flux as an additional parameter in our grid of 
photochemical models.  We choose stellar spectra that represent two bounding 
cases of a low and high UV flux as shown in Figure~\ref{f1}.  It is our 
assumption that the actual UV spectrum for GJ 1214 falls somewhere in between 
these two extremes.  For our low UV case, we employ a stellar model with the 
same $T_{eff}$ and $g_{surf}$ as GJ 1214, which we calculate by interpolating 
between bracketing models by \citet{hau99}.  The stellar model does not include
sources of UV emission due to stellar activity, so it produces a spectrum that 
drops off precipitously in the UV following the stellar blackbody, which is
almost certainly an underestimate for the actual UV flux from GJ 1214.  For our
high UV case, we take the observed time-averaged spectrum of the active M4.5V 
flare star AD Leo from \citet{seg05} ($T_{eff} = 3400$ K), which is one of the 
most active known M-stars.  The incident stellar spectrum is then calculated at
the orbital distance of GJ 1214b, from 100 to 1000 nm, and this flux is used 
for calculating photodissociation rates for the 33 photolysis reactions 
included in the photochemical kinetics code.

\subsection{Transmission Spectrum Modeling \label{tr_spec}}

Using the chemical abundances profiles that we obtain from the photochemical 
modeling, we calculate transmission spectra using the model outlined in 
\citet{mil09} for super-Earth exoplanets.  We calculate absorption of stellar 
light along chords through the planet's upper atmosphere and then integrate to 
determine the total absorption over the entire projected annulus of the 
planet's atmosphere in transit.  This results in a wavelength-dependent 
transmission spectrum, which we calculate from 0.3 - 30 $\mu$m.  As with the 
photochemical modeling, we employ the T-P profile from \citet{mil10} for a 
solar composition atmosphere when calculating the spectra.

To calculate the transmission spectra we include opacities for the major 
molecular absorbers in the optical and IR including H$_2$O 
\citep{fre08, par97}, CH$_4$ \citep{fre08, kar94, str93}, NH$_3$, CO 
\citep[][and references therein]{fre08}, and CO$_2$ \citep{rot05}.  We 
additionally include absorption from a number of non-equilibrium carbon-bearing
species -- HCN \citep{har08}, C$_2$H$_2$, C$_2$H$_4$, and C$_2$H$_6$ 
\citep{rot05}  -- which have significant IR cross sections and are predicted 
to be present at high abundances in some of our photochemical calculations.  We
caution that some of the opacity lists, especially those for C$_2$H$_2$, 
C$_2$H$_4$, and C$_2$H$_6$ along with the short wavelength CH$_4$ data may be 
incomplete.  Atomic species with large opacities in the optical -- Na and K -- 
are predicted to be condensed out of the atmosphere of GJ 1214b at the heights 
probed by transmission spectroscopy and are not included in our calculations.  
We do however include the effects of Rayleigh scattering in the optical for the
most abundant molecules in the atmosphere (H$_2$, He, H$_2$O, CH$_4$, NH$_3$, 
CO, CO$_2$, and N$_2$).  We calculate Rayleigh scattering cross sections for 
each molecule individually according to
\begin{equation}
\sigma = \frac{8\pi}{3}\left(\frac{2\pi}{\lambda}\right)^4\alpha^2
\end{equation}
where $\alpha$ is the polarizability obtained from the CRC Handbook for each
molecule.  We do not include the effects of additional scattering into or out 
of the beam or refraction, which has been found to have a minimal effect 
\citep{hub01}.  

The end result is a suite of 18 transmission spectra for possible chemical
compositions of GJ 1214b's atmosphere.  Spectra are calculated for three values
of metallicity -- 1, 5, and 30 times solar; 3 values of $K_{zz}$ -- $10^6$, 
$10^7$, and $10^9$ cm$^2$/s; and two stellar UV input spectra.  

\section{Results \label{results}}

\subsection{Photochemistry \label{results_chem}}

\begin{figure}
\plotone{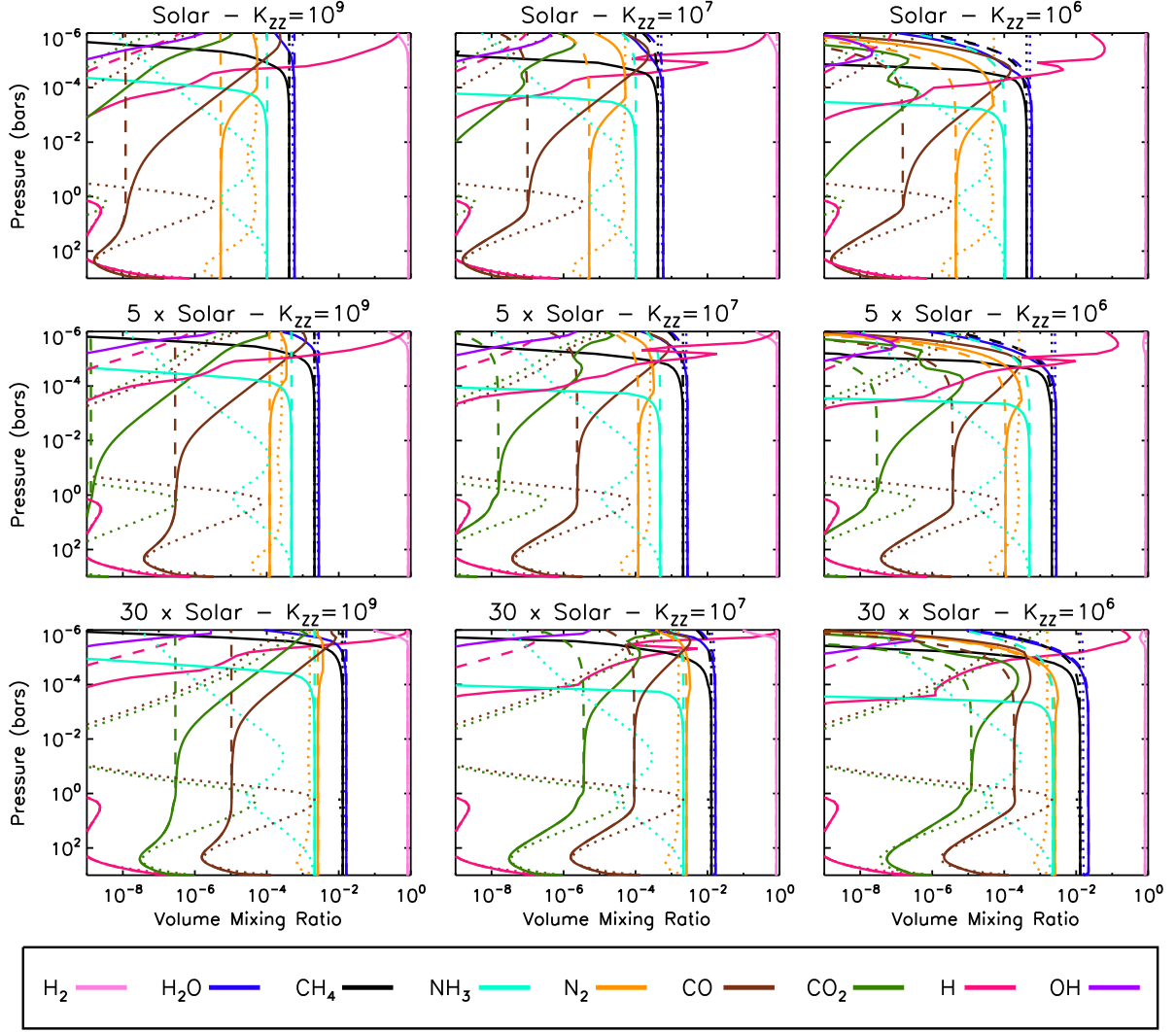}
%\begin{center}
%\includegraphics[angle=90.0, scale=0.73]{f2.eps}
%\end{center}
\caption{Results from our photochemical calculations for the major atmospheric
        constituents.  Models using the quiet M-star as the stellar input 
	spectrum are shown with dashed lines.  Models using AD Leo as the input
	spectrum are shown with solid lines.  Equilibrium abundances are 
	shown with dotted lines for reference.  	
        \label{f2}}
\end{figure}

\begin{figure}
\plotone{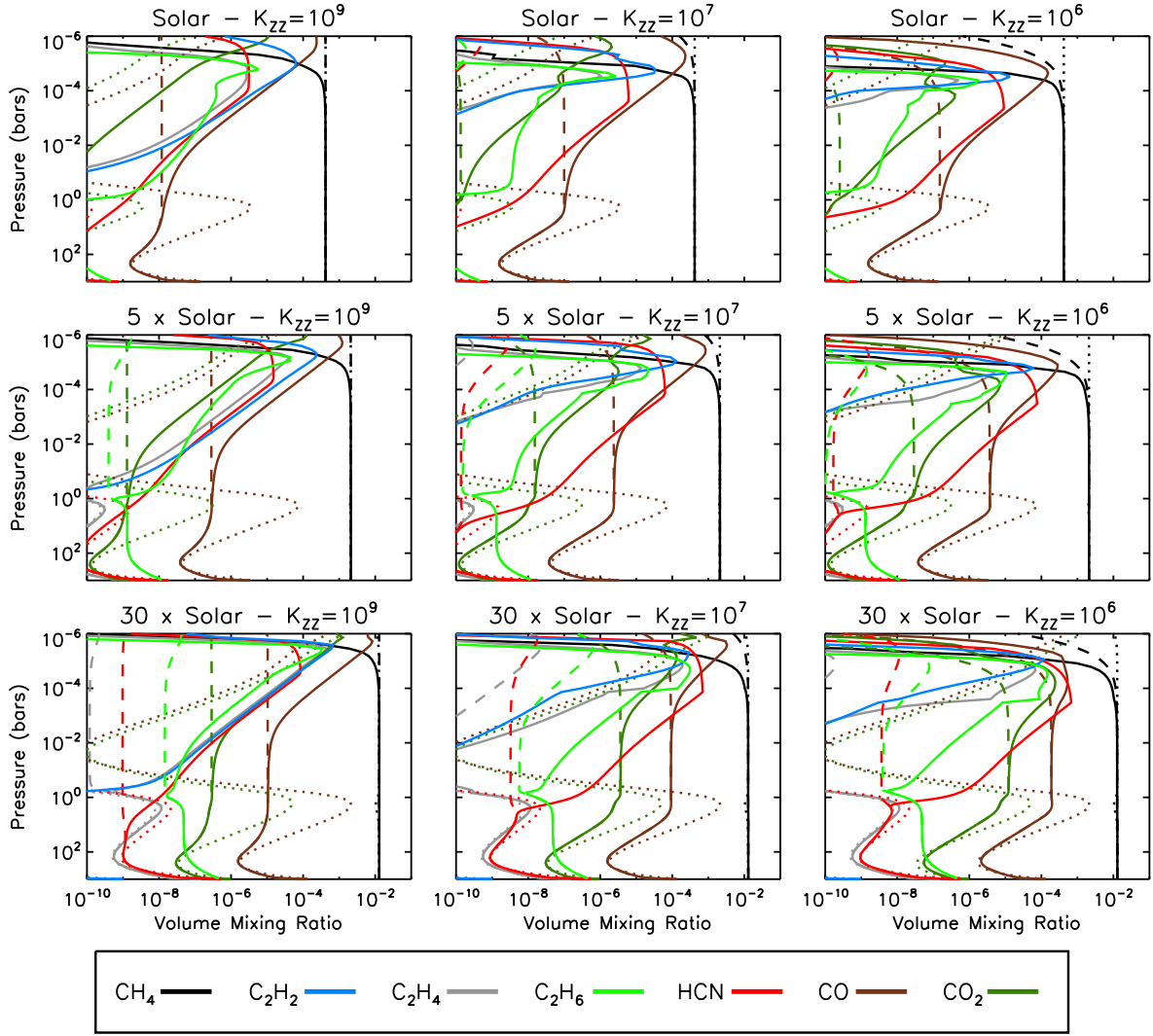}
\caption{Same as Figure~\ref{f2} but for the major carbon-bearing molecules 
        only.
        \label{f3}}
\end{figure}

Results from running the photochemical kinetics code with our suite of 18 
permutations on the model parameters are shown in Figure~\ref{f2} (major
atmospheric constituents) and Figure~\ref{f3} (carbon-bearing molecules).  
Depending on the input parameters, the resulting atmospheric abundance profiles
can differ dramatically from what is obtained from equilibrium chemistry.  This
is a result of both photochemistry, which breaks apart molecules in the upper 
atmosphere, and vertical mixing, which redistributes species throughout the 
atmosphere.  Qualitatively, the effect of eddy mixing is to smooth out chemical
gradients in the lower atmosphere, resulting in a more evenly distributed 
chemical composition.  In the upper atmosphere the effects of UV photolysis and
molecular diffusion dominate.  

Chemically, the main source of differences between the predictions of 
equilibrium chemistry and our photochemical modeling is that certain reactions
occur so slowly that the chemistry will never converge to equilibrium before
photolysis and mixing take over.  Specifically, photolysis of both ammonia and 
methane occur readily in the upper atmosphere, but reactions that create these 
molecules are not favored.  The end result is that nitrogen and carbon 
preferentially form into molecules other than methane and ammonia, resulting in
lower abundances of both of these species in the upper atmosphere than what is 
predicted by equilibrium calculations.  Nitrogen tends to combine into the very
stable N$_2$ molecule and also forms a smaller amount of HCN.  Carbon is 
redistributed into a wide variety of molecules including CO, CO$_2$, C$_2$H$_2$
(acetylene), C$_2$H$_4$ (ethylene), C$_2$H$_6$ (ethane), and HCN (hydrogen 
cyanide), resulting in a complex carbon chemistry  as shown in Figure~\ref{f3}.

In our models using the quiet M-star as the UV input spectrum (dashed lines in
Figures~\ref{f2} and \ref{f3}) very little UV photolysis takes place.  These 
results can therefore be interpreted as showing the effects of vertical mixing 
in the absence of any significant photochemistry.  In these models, the 
atmosphere is generally well-mixed, and abundance profiles for most major 
constituents are constant with height throughout most of the atmosphere until 
molecular diffusion takes over at high altitudes.  This is in sharp contrast to
the predictions of equilibrium chemistry that produce strong gradients in the 
abundances of several key molecules including CO, CO$_2$, and NH$_3$ (dotted 
lines in Figures~\ref{f2} and \ref{f3}).  Our model results produce obvious
quenched behavior for several molecules at modest depth including CO, CO$_2$ 
and HCN.  For these molecules, abundance profiles clearly follow their 
equilibrium values at depth, whereas their abundances are fairly constant above
the quench point.  All of these molecules also display a gradual transition 
between the quenched regime and the equilibrium regime, which is a result of 
atmospheric mixing.  In the absence of mixing, the abundance profiles would 
transition sharply from equilibrium to their constant quenched abundances at 
the quench point.  CO, CO$_2$, and HCN display fully quenched behavior above
1-10 bar.   In contrast, N$_2$ and NH$_3$  display quenched behavior even at 
the very base of the atmosphere for all of our models, which implies that the
actual quench point for these species lies deeper than 1000 bar.  For this 
reason, it is possible that the actual abundances of nitrogen-bearing species 
in GJ 1214b's atmosphere may differ somewhat from the values we report here.  

Without high levels of UV irradiation, methane remains the dominant 
carbon-bearing species throughout the atmosphere and is present at high 
abundances ranging from 0.1\% for solar metallicity atmospheres to 1\% when the
metallicity is enhanced to 30 $\times$ solar.  Ammonia and N$_2$ are expected 
to be the most abundant nitrogen-bearing molecules with N$_2$ becoming 
increasingly abundant at higher metallicities.  Some additional carbon-bearing 
species appear at moderate abundances for models with low $K_{zz}$ and high
 metallicity including HCN and C$_2$H$_6$.   Molecular diffusion allows for 
heavy molecules to preferentially settle out of the atmosphere at pressures 
lower than $\sim 100$ $\mu$bar for models with $K_{zz}$ of $10^6$ cm$^2$/s and 
$\sim 10$ $\mu$bar for models with $K_{zz}$ of $10^7$ cm$^2$/s.   

\begin{figure}
\plotone{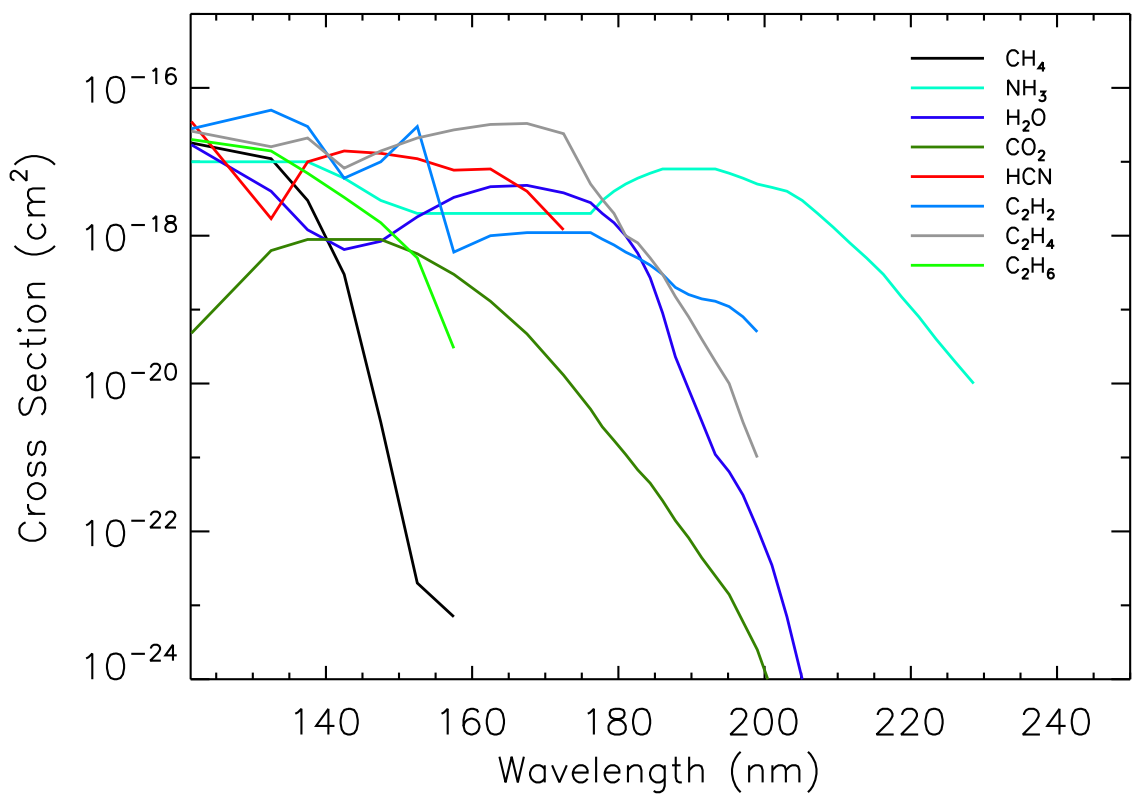}
\caption{Photolysis cross sections for some of the molecules that are 
        predicted to be present at high abundance in GJ 1214b's atmosphere.  
	Photolysis rates are detemined by 
	$\int \sigma F_{\nu} e^{-\tau_{\nu}}d\nu$ where $F_{\nu}$ is the
	stellar flux density, $\tau_{\nu}$ is the optical depth, and $\sigma$ 
	is the photolysis cross section.   The photolysis cross sections 
	plotted here are from \citet[][and references therein]{zah11}.
        \label{photolysis}}
\end{figure}

For the highly irradiated models, the upper atmosphere chemistry is further
complicated by UV photolysis.  In these atmospheres, the chemistry is driven by
photolysis of methane and ammonia.  Ammonia has an appreciable photolysis cross
section throughout the UV, making it unstable in the upper atmosphere 
(see Figure~\ref{photolysis}).  Methane only has a large UV cross section 
shortward of 1400 \AA, but it experiences significant photolysis from
Lyman-$\alpha$ photons at 1216 \AA.  The heights at which CH$_4$ and NH$_3$ 
are removed from the atmosphere has a strong dependence on the amount of 
vertical mixing.  At higher values of $K_{zz}$, methane and ammonia are lofted 
higher into the atmosphere, which results in a replenishing source that 
counteracts the effects of photolysis.  As the models increase in metallicity,
both methane and ammonia also mantain higher abundances to higher altitudes.  
This results from the overall higher abundance of each of these molecules 
at high Z.  Ammonia photodissociates at pressures ranging from 0.1 mbar
for high metallicity and high $K_{zz}$, to several mbar for solar metallicity 
and low $K_{zz}$.  Methane is 
stable to somewhat higher altitudes corresponding to pressures of $\sim$10
$\mu$bar for high metallicity and high $K_{zz}$, to $\sim$100 $\mu$bar for 
low metallicity and low $K_{zz}$.  Despite methane's propensity to remain 
stable even at fairly high altitudes in GJ 1214b's atmosphere, the combination 
of methane being converted into other carbon-bearing molecules at altitude 
along with downward transport produces a complex carbon chemistry throughout 
the atmosphere.  For all of the models with high UV irradiation, CO and CO$_2$ 
abundances steadily increase as a function of altitude until reaching the 
region of the atmosphere that is affected by molecular diffusion.  HCN, 
C$_2$H$_2$, C$_2$H$_4$, and C$_2$H$_6$ all demonstrate similar behavior with 
abundances increasing in the upper atmosphere due to photochemical production 
before tailing off again at the very top of the atmosphere due to molecular 
diffusion and UV photolysis.  The 
abundance gradients for each of these species is highly dependent on 
metallicity and vertical mixing.  At high metallicity, HCN, C$_2$H$_4$, and 
C$_2$H$_6$ achieve modest abundances even at the very base of the atmosphere.  
Here again the effect of mixing is to smooth out vertical chemical gradients, 
which results in the non-equilibrium carbon species becoming increasingly 
well-mixed at higher values of $K_{zz}$.

Aside from methane and ammonia, other molecular species such as H$_2$O, 
C$_2$H$_2$, C$_2$H$_4$, C$_2$H$_6$, and HCN experience significant UV 
photolysis in the upper atmosphere in all of our highly irradiated models.  As 
a result, the abundances of these molecules diminish rapidly at high altitude
($P \sim 10^{-6}$ bar), even for our models with high $K_{zz}$.  Conversely, 
for the highly irradiated high K$_{zz}$ models, photochemical 
\textit{production} of N$_2$, CO, and CO$_2$ leads to increasing abundances of 
these molecules at altitude given their negligible photolysis cross sections 
over the wavelength range where UV photons readily penetrate the atmosphere.  
Atomic species along with the OH radical also become increasingly 
abundant at high altitude as products of photolysis reactions.  At low 
values of $K_{zz}$, molecular diffusion plays a significant role high in the 
atmosphere, and in this case heavy molecules are not present at altitude due to
gravitational settling.  For these low $K_{zz}$ models, the upper atmosphere 
chemistry is driven by a combination of both molecular diffusion and 
photolysis.  

%Due to the complexity of the chemistry of GJ 1214b's atmosphere, as shown in 
%Figures~\ref{f2} and \ref{f3}, we outline our results for specific molecules
%below. CO-AUTHORS -- IS THIS NECESSARY?? 

%\subsubsection{NH$_3$ and N$_2$}

%\subsubsection{CH$_4$, CO, and CO$_2$}

%\subsubsection{H$_2$, H$_2$O, H, and OH}

%\subsubsection{C$_2$H$_2$, C$_2$H$_4$, and C$_2$H$_6$}

%\subsubsection{HCN}

\subsection{Transmission Spectroscopy}

\begin{figure}
\plotone{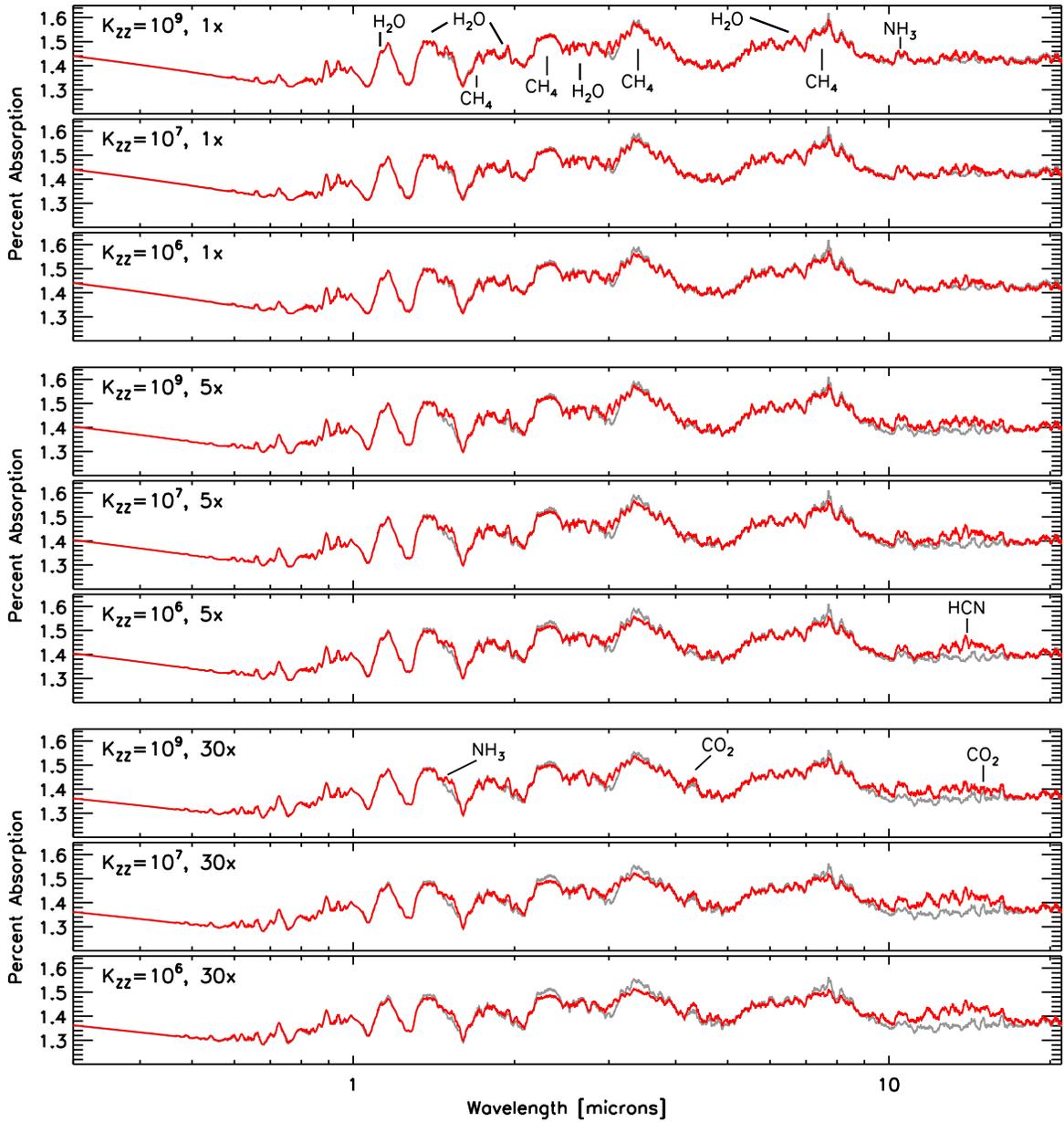}
\caption{Transmission spectra for GJ 1214b given the atmospheric composition 
        shown in Figures ~\ref{f2} and ~\ref{f3} for an active host star (AD 
	Leo input spectrum) - red lines.  For comparison, the spectra obtained 
	for atmospheres in thermal equilibrium are shown in gray.  For each 
	panel, values for $K_{zz}$ and metallicity are indicated.  Departures 
	from the equilibrium spectra are mostly due to the re-budgeting of 
	carbon away from methane into other molecules such as CO$_2$, and HCN. 
	Major spectral features are as indicated.  All spectra have been 
	normalized so as to reproduce the observed transit depth of 1.35\% in 
	the MEarth bandpass.
        \label{f5}}
\end{figure}

\begin{figure}
\plotone{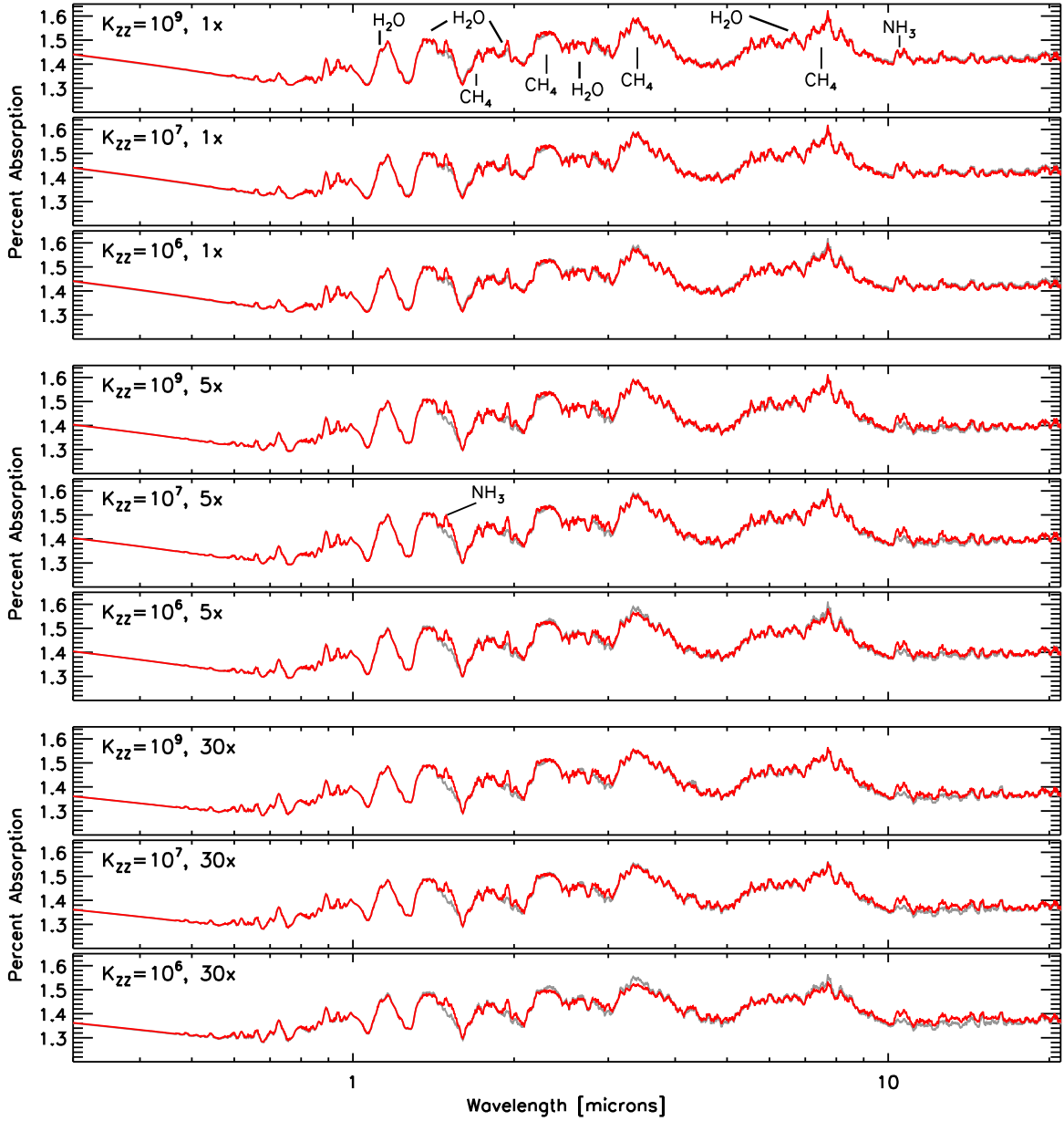}
\caption{Same as Figure ~\ref{f5}, but for a quiet host M-star.  Here the 
        non-equilibrium chemistry plays a lesser role in affecting the
	transmission spectrum, since methane remains stable to high elevations
	in the planet's atmosphere.
        \label{f6}}
\end{figure}

We take the abundance profiles from our photochemical modeling and use these
to produce theoretical transmission spectra for GJ 1214b for each of our 18 
models.  These results are shown in Figure~\ref{f5} for the highly irradiated
atmosphere and Figure~\ref{f6} for models using the quiet M-star as the input
stellar spectrum.  In these figures, all of the spectra have been normalized so
as to reproduce the observed transit depth of 1.35\% in the MEarth filter
bandpass spanning 675 to 1050 nm \citep{cha09}.  

For the low irradiation models, the spectra that we obtain look virtually 
identical to the spectra calculated using equilibrium chemistry (gray lines in
Figure~\ref{f6}).  For these models, photodissociation does not play a strong
role in perturbing the atmospheric chemistry.  While the overall atmospheric
chemistry is expected to be more well-mixed from our calculations than what is
obtained from equilibrium chemistry, there is very little effect on the species
that produce most of the opacity in our models.  Water, methane, ammonia, and
to a lesser extent CO and CO$_2$ all have significant opacities between 0.3 and
20 $\mu$m.  Of these, the spectral features of water and methane dominate the
transmission spectrum, and the abundances of these two species do not differ 
substantially from their predicted equilibrium values at the height probed by 
transmission spectroscopy (generally $\sim1$ mbar, but can vary between $
0.1$ and $500$ mbar).  The CO and CO$_2$ abundance profiles from 
Figures~\ref{f2} and \ref{f3} do differ dramatically from their equilibrium 
abundances above $\sim$100 bar, however this is difficult to discern from 
looking at the transmission spectra.  One reason for this is that the CO and 
CO$_2$ abundances remain low throughout all of our low-irradiation models with
CO and CO$_2$ abundances not exceeding $10^{-4}$ and $10^{-5}$ respectively.
Additionally most of the predicted CO spectral features overlap with stronger 
absoprtion features from H$_2$O and CH$_4$, which essentially swallow up the 
spectral signature of what would otherwise be an excellent tracer molecule for 
the presence of atmospheric mixing and photochemistry.  For ammonia, we tend to
predict somewhat higher abundances than what is obtained from equilibrium 
predictions, and this has a small effect in increasing the ammonia absorption 
in features at 1.5 and 10.5 $\mu$m.  However, we note that this result is 
somewhat dependent on the quenching behavior of ammonia at depth and may not be
accurate since our photochemical models to not extend down to the NH$_3$/N$_2$ 
quench point.  We also tend to see very slightly decreased methane absorption 
in some of the models, but once again, this is a very small effect and is 
unlikely to be observable.

Perhaps more surprising is the appearance of the spectra using the high-UV
irradiation models (Figure~\ref{f5}).  Despite markedly different overall 
atmospheric chemistry than what is predicted from equilibrium expectations, the
effect on the transmission spectra here is also fairly minimal.  Methane and 
ammonia photodissociation lead to slightly weaker spectral features for both of
these molecules.  Additionally, for models with high levels of CH$_4$ 
photolysis, HCN contributes significant opacity around 14 $\mu$m, and increased
abundances of CO$_2$ also add spectral features at 4.3 and 15 $\mu$m for the 
high metallicity models.  It has been previously pointed out that CO and CO$_2$
abundances increase strongly with metallicity \citep{zah11}.  We find the same
result here, and note that we see increasing absorption from CO$_2$ in 
the transmission spectra as a function of metallicity.  Generally, the 
abundance of CO$_2$ needs to exceed 1 part in $10^5$ for this molecule to be 
apparent in the transmission spectrum of GJ 1214b.  
Overall, while the changes in the transmission spectra relative to equilibrium 
expectations are noticeable, they are not dramatic.  The largest effects are 
seen at low $K_{zz}$ and high metallicity.  Given the current quality of 
observational transmission data, it is likely that it will be difficult to 
distinguish between the different possibilities with observations.  We explore 
this question further in Section~\ref{obs}  

As a note of caution, it is important to point out that transmission 
spectroscopy specifically probes the region of the atmosphere at the planet's 
day-night terminator (around the 1 mbar pressure level), since this is where 
all absorption of stellar light will take place.  It is possible that the 
terminator chemistry is more complex than what we have modeled here for a 
number of reasons.  Since GJ 1214b is expected to be tidally locked to its host
star, it is likely that chemical gradients exist between the day and night side
of the planet.  Observations of the planet in transmission may therefore be 
sensitive to the presence of longitudinal winds that carry material from the 
day side to the night side of the planet, or vice versa.  These winds could be 
an additional contributor to non-equilibrium chemistry that we are unable to 
address here using our 1-D model \citep[e.g.][]{coo06}.  3-D models of 
atmospheric dynamics on hot Jupiters show that strong winds that advect heat 
from the day side to the night side of the planet should be present at the 
heights probed by transmission spectroscopy at both the eastern and western 
terminators \citep{sho08, dob10, rau10}.  In this case the chemistry at the 
terminator should resemble that of the planet's day side, and our use of a 
day-side chemistry model is justified in calculating transmission spectra for 
GJ 1214b \citep[e.g.][]{for10}.  However, we caution that it is unclear how to 
scale the dynamical models for hot Jupiters to a super-Earth like GJ 1214b, and
therefore the atmospheric circulation for this planet is unconstrained at this 
point in time.  

\subsection{Clouds and Hazes}

Clouds would further complicate the appearance of the planet's transmission 
spectrum by efficiently scattering starlight at short wavelengths.  At longer 
wavelengths in the mid-IR, the spectrum of GJ 1214b should remain mostly 
unaffected by clouds.  Qualitatively, in the optical and near-IR, clouds can 
flatten the planet's transmission spectrum by scattering light at higher 
altitudes than where spectral features in transmission are expected to 
originate.  For GJ 1214b, we have previously found that clouds at pressures 
greater than 200 mbar can explain the observations of a flat transmission 
spectrum from 780 to 1000 nm by \citet{bea10}.  This value was determined by
cutting off transmission at different heights in the planet's atmosphere to 
simulate the effects of an optically thick gray cloud deck.  Clouds deeper in 
the atmosphere than 200 mbar would have only a minimal effect on the 
transmission spectrum at the wavelengths of the \citet{bea10} observations.  If
clouds are present, the exact shape of the transmission spectrum at short 
wavelengths is determined by whether the scattering is occurring within the 
Rayleigh or Mie regime, which will produce different power law slopes for the 
scattering opacity.  Unfortunately, self-consistently modeling the cloud 
opacity requires knowledge of the cloud particle size and height distributions,
which are currently unknown for GJ 1214b.  Here we offer up some suggestions 
for what the cloud composition could be.

\begin{figure}
%\begin{center}
\plotone{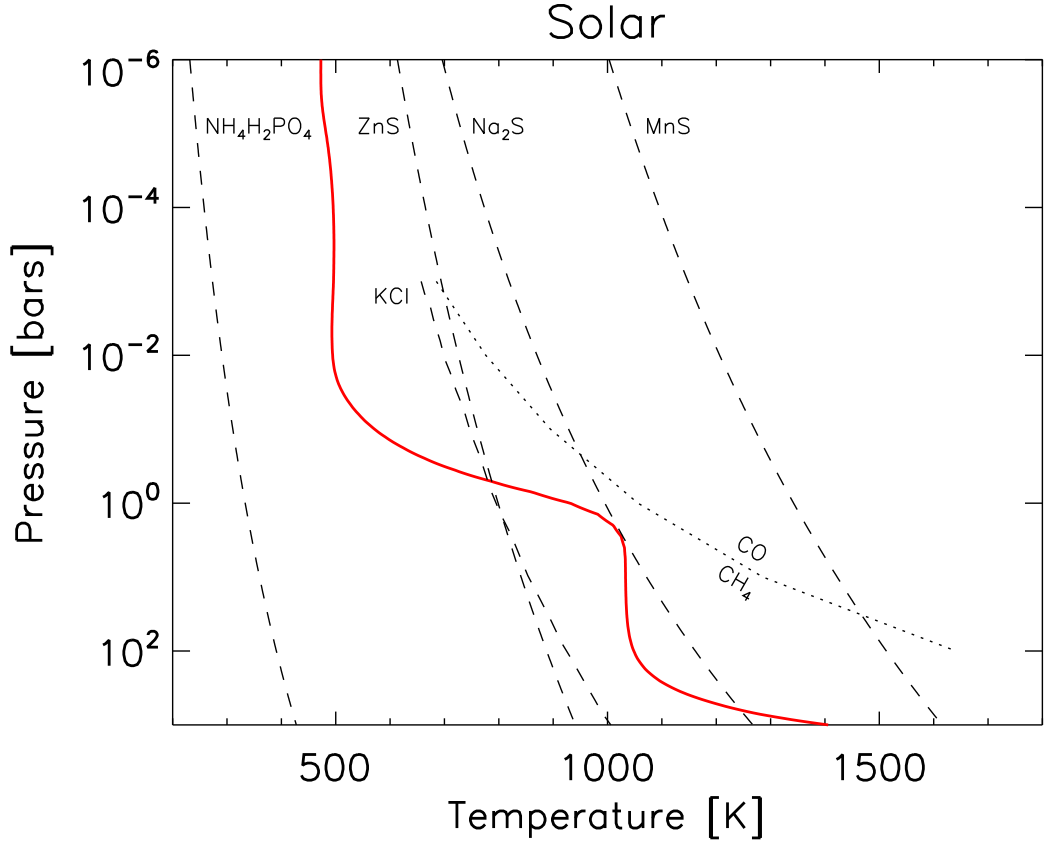}
%\includegraphics[scale=0.78]{f7.eps}
%\end{center}
\caption{The calculated T-P profile for GJ 1214b at solar metallicity.  
        Condensation curves for various molecules are overplotted as indicated.
	Species whose condensation curves cross the T-P profile will condense 
	in GJ 1214b's atmosphere at the pressure indicated.  KCl and ZnS are 
	predicted to both condense at 500 mbar for solar composition.  The 
	CO/CH$_4$ equilibrium curve is also shown (dotted line), indicated that
	methane is predicted to be the dominant equilibrium carbon-bearing 
	species throughout GJ 1214b's atmosphere.  The CO-CH$_4$ curve shifts 
	towards the left (to lower temperature) at higher metallicity and can 
	even intersect the T-P profile at 30-50 times solar metallicity.  
	However, this occurs above the quench point of CO/CH$_4$, so it has 
	little effect on the atmospheric chemistry.  
        \label{f7}}
\end{figure}

Clouds are formed by condensation processes, which will occur if the partial 
pressure of a species surpasses its vapor or condensation pressure.  To 
determine whether cloud formation will occur in the atmosphere of GJ 1214b, we 
compare the planet's T-P profile against the condensation curves of various 
molecules that are predicted to condense in hot planet and cool star 
atmospheres from \citet{lod06}.  The only molecules that we find whose 
condensation curves intersect the predicted T-P profile of GJ 1214b are KCl and
ZnS, as shown in Figure~\ref{f7}, although we caution that the T-P profile of 
GJ 1214b is unconstrained by observations, so its exact shape is somewhat 
uncertain.  If the T-P profile from Figure~\ref{f7} is correct, then both KCl 
and ZnS should condense at pressures of $\sim$500 mbar in GJ 1214b's 
atmosphere at solar composition.  For higher  metallicities a number of effects
come into play which can shift the condensation to either higher or lower 
pressure in GJ 1214b's atmosphere.  These effects are (1) the abundaces of 
condensate materials tend to be higher at higher Z, (2) the condensation 
curves for condensate species tend to shift to higher temperature i.e. to the 
right on Figure~\ref{f7} due to the increased vapor pressure of the heavy 
elements in the gas phase, and (3) the T-P profile tends to shift  up and to 
right on Figure~\ref{f7} due to the increased opacities from higher metal
abundances at high metalicity.  The first two of these effects will push 
condensation to higher pressures where a deeper and more massive cloud layer
will form.  The third effect (shifting the T-P profile) will have the opposite
effect of pushing condensation to higher in the atmosphere at lower pressure.  
KCl and ZnS are both predicted to be 
present at very low abundances under the assumption of equilibrium, so it is 
unclear if either of these species could be present in large enough quantities 
to form an optically thick cloud.  The predicted equilibrium abundance for KCl 
at 500~mbar and 800~K is only $8.5 \times 10^{-9}$ at solar metallicity and 
$1.5 \times 10^{-6}$ at 30 $\times$ solar metallicity.  We do not calculate 
abundances for ZnS in our equilibrium chemistry code since we only include the 
22 most abundant atomic elements from the Sun.  Zinc's abundance in the Sun is 
only $1.5 \times 10^{-8}$ \citep{asp05}, and ZnS is likely to be present in 
even lower amounts in GJ 1214b's atmosphere.  However, due to the fact that
GJ 1214b is viewed in transmission along a slant optical path, even low 
abundances of condensate particles can form an optically thick cloud.  
\citet{for05b} explored this issue and found that condensed KCl may form a 
cloud with an optical depth of $\sim$0.18 in a solar composition hot Jupiter 
atmosphere, whereas ZnS should only form a cloud of optical depth $\sim$0.09.

Another possibility is that GJ 1214b may form photochemical hazes in its 
atmosphere.  The composition of any such hazes is unconstrained due to the 
complex nature of the photochemical processes that would lead to their 
formation.  However, we can look to solar system objects to inform our 
understanding of possible haze formation mechanisms for GJ 1214b.  
Photochemically induced hazes are common in the atmospheres of solar system 
planets and planet-sized moons.  Among the nearby planets, Venus possesses a 
sulfuric acid haze layer in its atmosphere.  Additionally, a number of solar 
system objects including Jupiter, Neptune, and Titan have hazes composed of 
complex hydrocarbons.  Hazes may be common in exoplanet atmospheres as well.  
The transiting hot Jupiter HD 189733b has an optical transmission spectrum that
is consistent with a high altitude haze \citep{sin11, pon08}, although its 
composition is currently unconstrained.  Of the solar system haze scenarios, it
is most likely that GJ 1214b forms hydrocarbon hazes in its atmosphere, given 
the high abundance of methane that we predict.  Many of our photochemical 
models from Section~\ref{results_chem} produce large amounts of the 
second-order hydrocarbons C$_2$H$_2$, C$_2$H$_4$, and C$_2$H$_6$, which are the
first byproducts in the sequence of chemical reactions to form the high-level 
hydrocarbons that would make up a Titan-like haze.  Unsaturated hydrocarbons 
such as C$_2$H$_2$ have a propensity to polymerize, eventually forming complex 
molecules such as PAH's, tholins, and soots \citep[e.g.][]{yun84}.  
Unfortunately, our photochemical code does not include reactions for 
hydrocarbons of higher order than C$_2$H$_X$, so we cannot predict the exact 
composition or abundance profiles for the haze layer.  We can however state 
that GJ 1214b should be amenable to forming hydrocarbon haze in its 
atmosphere. 

\section{Confronting the Observations \label{obs}}

A number of observational studies of GJ 1214b's atmosphere have already taken
place, and here we attempt to rectify our modeling efforts with the available 
data.  To date, observations of GJ 1214b's atmosphere have been focused 
on transit measurements to determine the planet's effective radius as a 
function of wavelength (transmission spectrum).  An alternative observing
strategy would be to observe GJ 1214b at secondary eclipse to obtain day-side 
emission spectroscopy.  Unfortunately, the expected secondary eclipse 
depth for GJ 1214b at wavelengths shortward of 5 $\mu$m is less than 100 ppm 
\citep{mil10} and is therefore likely to remain beyond the reach of 
observational detection for current instrumentation.  In the meantime, 
transmission spectroscopy has the added benefit of being highly sensitive to 
the atmospheric mean molecular weight, which allows for a better understanding
of the bulk composition of the atmosphere.   

The observational efforts to date are summarized as follows.  The MEarth survey
produced the original measurement of GJ 1214b's transit depth using a 
broad-band optical/near-IR filter spanning $\sim$675-1050 nm \citep{cha09}.  
These transit observations produced a transit depth of $1.35$\%, which was 
later refined by \citet{ber11} to $1.37$\% using additional epochs of MEarth 
observations along with transit observations from the VLT and FLWO 1.2-meter 
telescope.  \citet{kun10} further extended the transit observations to shorter 
wavelengths at \textit{r}-band, obtaining a transit depth of 
$(R_{p}/R_{\star})^2 = 0.01084 \pm 0.00056$.  However, their $r$-band data is
highly degenerate with limb darkening, and including the formal error bars on
their limb darkening parameters leads to a much larger error bar in the transit
depth.  \citet{bea10} produced the first wavelength-dependent
transmission spectrum in the very-near-IR from 780 to 1000 nm, using the
FORS2 instrument on the VLT.  The resulting transmission spectrum is 
featureless, as discussed in Section~\ref{intro}.  \citet{des11} provide two
additional data points to the transmission spectrum at longer wavelengths in 
the IR using the two warm Spitzer IRAC channels.  Their 3.6 and 4.5 $\mu$m data
continue to reveal a flat transmission spectrum at high significance.  
\citet{cro11} are the only authors to report a wavelength dependent 
\textit{variation} in transit depth for GJ 1214b.  Their J and Ks band data 
show a strong discrepancy between the transit depths observed at these two 
wavelengths at a level of several sigma.  This is attributed to GJ 1214b having
a hydrogen-rich atmosphere with a large scale height, which is the only way to 
produce such a large difference between the transit depths observed at two 
separate wavelengths.

Here we attempt to fit all the available data with our modeled transmission
spectra from Section~\ref{tr_spec}.  We compare each of our modeled spectra to
the observed transit depths from all of the data sets where multi-wavelength 
data was obtained, and we compute a goodness-of-fit ($\chi^2$) parameter to 
assess how well each model reproduces the data.  We perform a comparison to the
ensemble of multi-wavelength data from \citet{bea10}, \citet{des11}, and 
\citet{cro11}.  We additionally do a comparison between our models and only the
combined \citet{des11}, and \citet{cro11} data, under the assumption that the 
\citet{bea10} data at wavelengths shortward of 1 $\mu$m are affected by the 
presence of clouds.  We normalize all of our models vertically such that the 
best fit to the data is achieved, and we do not allow for vertical offsets 
between the individual datasets, which would introduce an unnecessarily large 
number of free parameters into our analysis.  While GJ 1214 is known to be an 
active and spotted star, the wavelength-dependent effect on the transit depth 
should be minimal and has already been accounted for in the observational 
studies.

We compare the data against a total of 27 different transmission spectrum 
models that we have generated (listed in Table~\ref{t1}).  The first 18 are the
results of our photochemical modeling, shown in Figures~\ref{f5} and \ref{f6}. 
Four of the remaining models are for hydrogen-rich atmospheres -- three are 
equilibrium compositions of 1, 30, and 50 times solar metallicity from 
\citet{mil10}, and the fourth is a solar composition (equilibrium) atmosphere 
with methane artificially removed simulating the effect of much stronger UV 
photolysis than what we have reported in Section~\ref{results_chem}.  The 
remaining five models are for atmospheres with increasing mean molecular weight
due to higher abundances of water relative to H$_2$.  We calculate models 
consisting of 10, 20, 30, 40, and 100\% H$_2$O, with the remaining portion of 
the atmosphere composed of H$_2$.

\begin{figure}
%\plotone{f8.eps}
\begin{center}
\includegraphics[angle=90.0, scale=0.68]{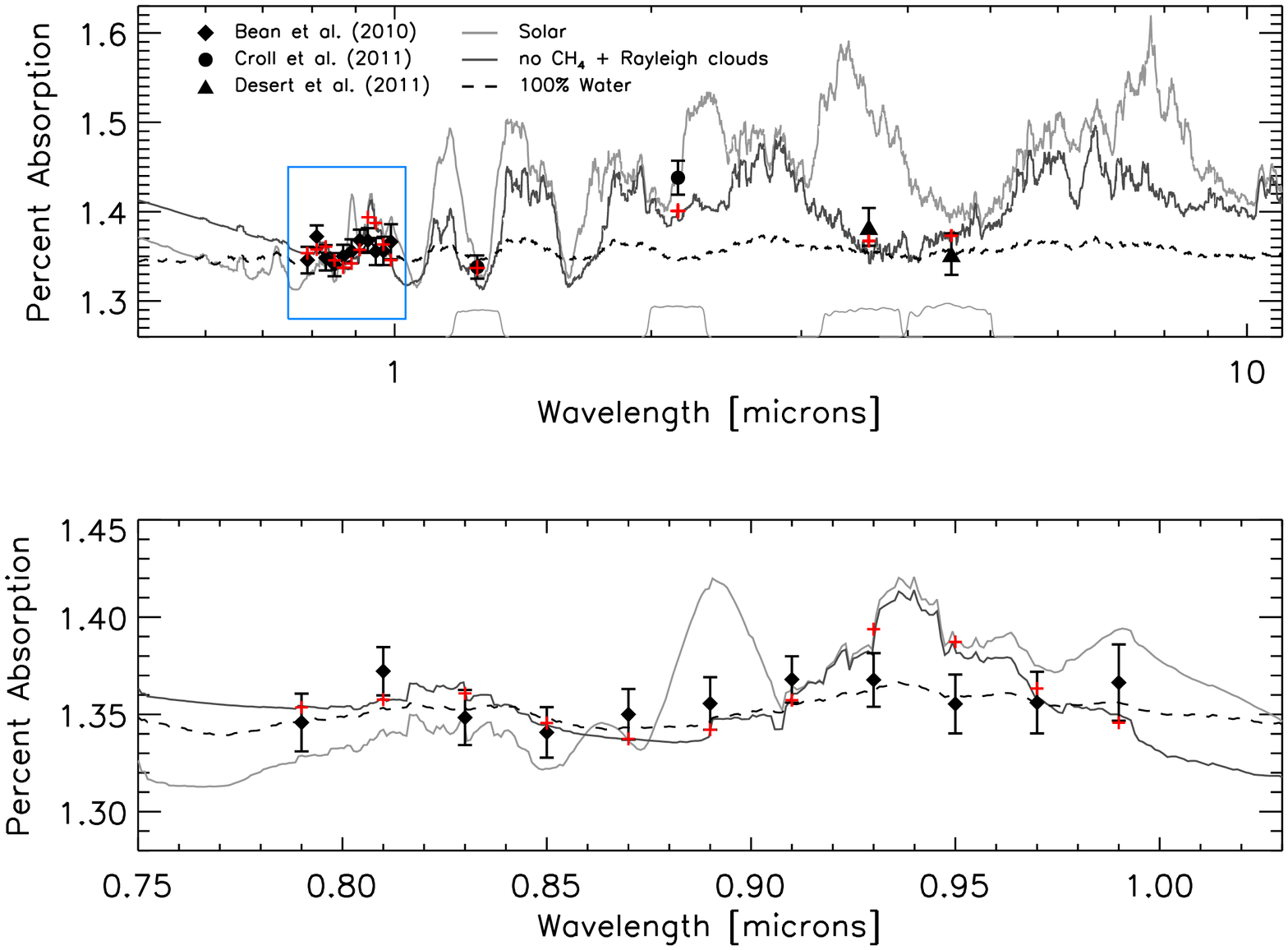}
\end{center}
\caption{Comparisons of the transmission data for GJ 1214b from \citet{bea10}, 
        \citet{des11}, and \citet{cro11} to our model spectra (top).  The same
	is shown for just the \citet{bea10} data in the bottom panel.  The 
	bandpasses for the \citet{des11} and \citet{cro11} data are shown at
	the bottom of the upper panel for reference.  The best fit to all of 
	the data is obtained for a solar composition atmosphere with methane 
	completely removed and clouds affecting the spectrum at short 
	wavelengths (dark gray solid curve).  Clouds have been simulated 
	here by increasing the nominal Rayleigh scattering opacity by a factor 
	of 5.  Red ``$+$'' signs show the expected values of the transit depth
	based on the best-fit model, averaged over the bandpasses of each
	observed data point.  The transmission spectrum for a 100\% water steam
	atmosphere provides the second-best fit to the data (dashed black 
	curve) but is in strong disagreement with the \citet{cro11} Ks-band 
	data point.  The transmission spectrum for a solar composition 
	atmosphere in chemical equilibrium is overplotted for reference (solid 
	light gray curve).  
        \label{f8}}
\end{figure}

For the combination of all of the available spectral data from \citet{bea10}, 
\citet{des11}, and \citet{cro11}, the best fit is achieved for an atmosphere
composed of 100\% water vapor.  The quality of the fit is not very good, 
remaining consistent with the ensemble of data at only the 2-$\sigma$ level.  
Here the fit to the model is dominated by the 11 data points from 
\citet{bea10}, which are an excellent match to the flat spectrum produced by a 
high mean molecular weight water atmosphere.  However, the \citet{cro11} 
Ks-band data point remains highly inconsistent with the steam atmosphere model 
at more than 5-$\sigma$ (see Figure~\ref{f8}).   It is tempting to consider the
\citet{cro11} Ks-band point as a statistical outlier under the steam atmosphere
scenario.  However, we point out that the \citet{cro11} results are robust in 
that three separate transits were observed in their study, and during all three
events the Ks band transit was observed to be deeper than the J-band transit by
a comparable magnitude.  

Given the poor model fits to the combination of all of the available 
transmission data, we consider the possibility that the \citet{bea10} results 
are influenced by clouds, but that the longer wavelength data remain 
unaffected.  Fitting the \citet{des11} and \citet{cro11} data alone, we find 
that the best fit is achieved for a solar composition atmosphere with methane 
artificially removed.  The quality of the fit is quite good and is consistent 
with the data at 1-$\sigma$.  The no-methane interpretation is intriguing in 
that it requires methane to be absent from the atmosphere at pressures of 
$\sim$1 mbar, where the transmission spectrum originates.  This directly 
contradicts the results of our photochemical modeling from 
Section~\ref{results_chem} where we found that methane should only photolyze to
pressures of less than $\sim$0.1 mbar and below this height it should be quite 
abundant.  One possibility to rectify this discrepancy is that GJ 1214b's 
atmosphere has an eddy diffusion rate of less than $10^6$ at altitude, which 
would allow for methane to be removed from the atmosphere at lower altitudes 
and higher pressures.  Another possibility is that we have underestimated the 
UV flux incident on the planet's upper atmosphere, although this is unlikely 
given that AD Leo is one of the most active known M-stars, and we used its UV 
spectrum for our high-irradiation models.  

The no-CH$_4$ model provides a poor fit to all of the available data when the
\citet{bea10} data is added in.  However, if we assume that clouds affect
the short-wavelength data, then this is to be expected.  To simulate the effect
of clouds in the wavelength range of the \citet{bea10} data, we increase the
amount of Rayleigh scattering by a factor of 5, which raises the height at 
which the atmosphere becomes optically thick from scattering.  We find that 
this provides a much improved fit to the data (see Figure~\ref{f8}).  The 
no-CH$_4$ atmosphere with increased Rayleigh scattering is consistent with
the combination of the \citet{bea10}, \citet{des11}, and \citet{cro11} at a
level of 1.5-$\sigma$.  The \citet{bea10} data alone similarly gives a 
1.5-$\sigma$ fit to the no-CH$_4$ increased Rayleigh scattering model.  
Even with the increased Rayleigh scattering, a water feature is expected to be
present in the transmission spectrum at 970 nm, which is absent from the
\citet{bea10} data, but the discrepancy here is still less than only 
2-$\sigma$.

If clouds are indeed the explanation for the short wavelength transmission 
data, the exact form of cloud scattering would differ from a simple 
$\lambda^{-4}$ power law for Rayleigh scattering.  Scattering occurs in the 
Rayleigh regime only if the particle size in the cloud layer is much smaller 
than the wavelength of the light.  For larger particle sizes approaching 1 
$\mu$m Mie scattering is expected, which should produce a shallower slope, more
closely approximating a gray cloud deck.  However, comparisons to forms of 
cloud opacity other than those approximated by Rayleigh scattering are beyond 
the scope of this work.  Observations that extend to shorter wavelengths and 
also filling in the data gap between 1 and 1.5 $\mu$m should help to further 
constrain the power law slope on the scattering opacity which could in turn 
provide hints as to the composition of the cloud deck.  

\section{Conclusions and Discussion \label{concl}}

While photochemistry should play an important role in determining the 
atmospheric chemistry of GJ 1214b, we find that its effects are minimal at the
heights probed by transmission spectroscopy.  We find only minor 
differences between our modeled transmission spectra for equilibrium chemistry
and the spectra resulting from our more detailed photochemical modeling.  
Methane and ammonia photolysis are expected to alter the atmospheric chemistry
by allowing additional carbon- and nitrogen-bearing molecules to form, some of
which do not appear in significant abundances under the assumption of chemical
equilibrium.  However, the effects of non-equilibrium chemistry do not extend
to pressures greater than 1 mbar at a sufficient level to have any major 
effects on our modeled transmission spectra.  

When fitting the available transmission spectroscopy data for GJ 1214b however,
an intriguing possibility is raised that methane photolysis may be much more
efficient than what we predict from our photochemical modeling.  The 
transmission data are best fit by a model with no methane and clouds or hazes
that affect the optical and near-IR spectrum.  This is consistent with GJ 1214b
having a hydrogen-rich atmosphere where methane is efficiently photolyzed, and
carbon-rich hazes form readily.  

Still, much degeneracy remains in the modeling efforts, which is due partly to 
the small amount of spectral data available for GJ 1214b's atmosphere.  
Additional data will provide much-needed constraints to our modeling efforts.  
It still remains to be confirmed that GJ 1214b does indeed possess a low mean
molecular weight atmosphere.  This interpretation is so far based upon a single
Ks-band data point from \citet{cro11}, which should be corroborated with 
additional data.  If methane is absent from GJ 1214b's atmosphere, then 
confirmation of GJ 1214b's low mean molecular weight through the observation
of deep spectral features in transmission should be done by observing water
features in the planet's spectrum at wavelengths longward of 1-2 $\mu$m where 
the effects of clouds should not be significant.  Additional observations at 
the wavelengths of methane features are also needed to confirm that this 
molecule is truly absent from the transmission spectrum.  The interpretation of
a low methane abundance so far depends strongly on the flat transmission 
spectrum observed by \citet{des11} with warm Spitzer, who should have seen 
strong methane absorption in the 3.6 $\mu$m IRAC band if this molecule was 
present in modest abundances.  

If GJ 1214b's atmosphere does in fact have a high mean molecular weight, then
the atmosphere must be water-rich to be consistent with models of the interior
of the planet \citep[e.g.][]{rog10}.  In this case the photochemistry should
be driven by photolysis of water into OH and H.  Such an atmosphere would 
surely be affected by escape of atomic hydrogen, resulting in the atmosphere
slowly becoming more oxidized with time.  The level of oxidation of the 
atmosphere should then be highly dependent on the rate of atmospheric escape, 
which is currently unconstrained.  Additional study of the photochemistry
of a water-rich atmosphere for GJ 1214b is beyond the scope of this work but
merits additional investigation.  
 
%It is important to point out that photochemical modeling is ultimately limited 
%by our knowledge of reaction rates and photolysis cross sections.  The reaction
%rates that we use in this paper are informed by both laboratory and theoretical
%work.  Often however, the full temperature-pressure dependence of a given 
%reaction rate or cross section is unknown.  Furthermore, most laboratory 
%experiments take place at far lower temperatures (and often higher pressures) 
%than what we expect in the upper atmosphere of a highly irradiated transiting 
%planet.  It is additionally possible that the full network of reactions for a 
%given molecule is poorly known, resulting in missing reactions in our current 
%table of 698 reaction rates.  Photochemical modeling, such as the work 
%presented in this paper has great merit in that it accounts for important 
%processes such as UV photolysis and atmospheric mixing that cannot be ignored 
%when considering atmospheric chemistry.  However, more accurate modeling will 
%ultimately require additional laboratory studies of reaction rates, especially 
%in the temperature-pressure regime that is relevant for exoplanet atmospheres. 

GJ 1214b is now one of a growing group of transiting super-Earth exoplanets.  
To date, 8 transiting super-Earths have been confirmed, and another 288 
``super-Earth'' candidates from Kepler with radii between 1.25 and 2 
$R_{\earth}$ still await confirmation \citep{bor11}.  Additionally, some of the
662 ``Neptune-like'' Kepler candidates with radii between 2 and 6 $R_{\earth}$
may also fit into the super-Earth category, when defined as in this paper by 
planets with masses between 1 and 10 $M_{\earth}$.  Interestingly, GJ 1214b 
falls into the latter category of planets that would be defined as 
``Neptune-class'' by the Kepler standards.  Looking at the current population 
of 8 transiting super-Earths, there seems to be a wide range of implied bulk 
composition of these planets, ranging from the iron-rich planet Kepler-10b 
\citep{bat11} to the hydrogen-rich planet Kepler-11e \citep{lis11}.  GJ 1214b 
fits in somewhere in the middle of this population of planets, with its 
middle-of-the-road density of 1.9 g/cm$^3$ and implied composition that is 
either ice-rich or hydrogen-rich (or somewhere in between these two extremes).
Ultimately, a combination of mass and radius measurements along with 
atmospheric studies such as the one presented here will allow for stronger 
constraints to be placed on the bulk composition of these planets, which will 
in turn give us a more detailed understanding of the formation and evolution 
processes that give rise to the observed population of super-Earths.

\acknowledgements 
E.~M.-R.~K was supported by a contract with the California Institute of 
Technology funded by NASA through the Sagan Fellowship Program.

\bibliography{ms}

\begin{deluxetable}{rl}
\tablecaption{Transmission Spectra Models\label{t1}}
\tablewidth{0pt}
\startdata
1 & Solar Composition\\
2 & 30 $\times$ Solar Composition\\
3 & 50 $\times$ Solar Composition\\
4 & Solar Composition w/ no CH$_4$\\
5 & 10\% H$_2$O, 90\% H$_2$\\
6 & 20\% H$_2$O, 80\% H$_2$\\
7 & 30\% H$_2$O, 70\% H$_2$\\
8 & 40\% H$_2$O, 60\% H$_2$\\
9 & 100\% H$_2$O (steam)\\
10 & Photochemistry -- 1 $\times$ Solar, K$_{zz}$ = 10$^9$ cm$^2$/s, Quiet M-star\\
11 & Photochemistry -- 1 $\times$ Solar, K$_{zz}$ = 10$^7$ cm$^2$/s, Quiet M-star\\
12 & Photochemistry -- 1 $\times$ Solar, K$_{zz}$ = 10$^6$ cm$^2$/s, Quiet M-star\\
13 & Photochemistry -- 5 $\times$ Solar, K$_{zz}$ = 10$^9$ cm$^2$/s, Quiet M-star\\
14 & Photochemistry -- 5 $\times$ Solar, K$_{zz}$ = 10$^7$ cm$^2$/s, Quiet M-star\\
15 & Photochemistry -- 5 $\times$ Solar, K$_{zz}$ = 10$^6$ cm$^2$/s, Quiet M-star\\
16 & Photochemistry -- 30 $\times$ Solar, K$_{zz}$ = 10$^9$ cm$^2$/s, Quiet M-star\\
17 & Photochemistry -- 30 $\times$ Solar, K$_{zz}$ = 10$^7$ cm$^2$/s, Quiet M-star\\
18 & Photochemistry -- 30 $\times$ Solar, K$_{zz}$ = 10$^6$ cm$^2$/s, Quiet M-star\\
19 & Photochemistry -- 1 $\times$ Solar, K$_{zz}$ = 10$^9$ cm$^2$/s, AD Leo (Active Star)\\
20 & Photochemistry -- 1 $\times$ Solar, K$_{zz}$ = 10$^7$ cm$^2$/s, AD Leo (Active Star)\\
21 & Photochemistry -- 1 $\times$ Solar, K$_{zz}$ = 10$^6$ cm$^2$/s, AD Leo (Active Star)\\
22 & Photochemistry -- 5 $\times$ Solar, K$_{zz}$ = 10$^9$ cm$^2$/s, AD Leo (Active Star)\\
23 & Photochemistry -- 5 $\times$ Solar, K$_{zz}$ = 10$^7$ cm$^2$/s, AD Leo (Active Star)\\
24 & Photochemistry -- 5 $\times$ Solar, K$_{zz}$ = 10$^6$ cm$^2$/s, AD Leo (Active Star)\\
25 & Photochemistry -- 30 $\times$ Solar, K$_{zz}$ = 10$^9$ cm$^2$/s, AD Leo (Active Star)\\
26 & Photochemistry -- 30 $\times$ Solar, K$_{zz}$ = 10$^7$ cm$^2$/s, AD Leo (Active Star)\\
27 & Photochemistry -- 30 $\times$ Solar, K$_{zz}$ = 10$^6$ cm$^2$/s, AD Leo (Active Star)\\

\label{composition}
\enddata
\end{deluxetable}

\end{document}